# Correlation between Magnetic Properties and Depinning Field in Field-Driven Domain Wall Dynamics in GdFeCo Ferrimagnets

Tomoe Nishimura[1], Duck-Ho Kim[1]★, Yuushou Hirata[1], Takaya Okuno[1], Yasuhiro Futakawa[2], Hiroki Yoshikawa[2], Arata Tsukamoto[2], Yoichi Shiota[1], Takahiro Moriyama[1], and Teruo Ono[1,3]★

[1]Institute for Chemical Research, Kyoto University, Uji, Kyoto 611-0011, Japan.

[2]College of Science and Technology, Nihon University, Funabashi, Chiba 274-8501, Japan.

[3]Center for Spintronics Research Network (CSRN), Graduate School of Engineering Science, Osaka University, Osaka 560-8531, Japan.

★Correspondence to: kim.duckho.23z@st.kyoto-u.ac.jp, ono@scl.kyoto-u.ac.jp

**Abstract**

The influence of temperature on the magnetic-field-driven domain wall (DW) motion is investigated in GdFeCo ferrimagnets with perpendicular magnetic anisotropy (PMA). We find that the depinning field strongly depends on temperature. Moreover, it is also found that the saturation magnetization exhibits a similar dependence on temperature to that of depinning field. From the creep-scaling criticality, a simple relation between the depinning field and the properties of PMA is clearly identified theoretically as well as experimentally. Our findings open a way for a better understanding how the magnetic properties influence on the depinning field in magnetic system and would be valuably extended to depinning studies in other system.

The dynamics of the domain walls (DWs) in magnetic materials has been extensively explored for understanding the physics [1–15] as well as for the potential applications in spintronic devices [16–20]. How the DW motion depends on the driving forces, such as magnetic field and electric current, brings up important fundamental questions [1, 3–15, 21–25]. In the absence of disorder (or for a large driving forces), dissipative viscous flow motion is dominant, of which DW velocity is proportional to the driving forces [4, 12–15, 26, 27]. In the real materials, however, local defects or inhomogeneities induce the presence of disorder leading to pinning. Pinning is known to have a strong effect on DW motion such as introducing stochasticity [28, 29] and inducing a DW roughness [1, 10]. A fundamental understanding on how pinning affects the DWs dynamics is thus important for applications. Although remarkable efforts have been made theoretically and experimentally [4, 12–15], it has remained elusive what are the key factors influencing the DW depinning. In this work, we address the question by providing a simple relationship between depinning field and magnetic properties.

In this study, we prepared GdFeCo ferrimagnetic films with perpendicular magnetic anisotropy (PMA) using magnetron sputtering. The detailed structures is 5-nm SiN/30-nm $Gd_{23.5}Fe_{66.9}Co_{9.6}$/100-nm SiN on intrinsic Si substrate. The GdFeCo films were then patterned into 5-μm-wide and 500-μm-long microwires with a hall cross structure using electron beam lithography and Ar ion milling. For current injection, 100-nm Au/5-mn Ti electrodes were stacked on the wire. To make an Ohmic contact, we removed the SiN capping layer using weak ion milling before electrode deposition. Figure 1shows schematic illustration of the device.

The field-driven DW velocity was examined by using a real time DW detection technique [21, 30, 31] as following this procedure. A large out-of-plane magnetic field (−200 mT) was applied to saturate magnetization and then the out-of-plane magnetic field $\mu_0 H$ in the opposite direction was applied for driving DW. Here, $\mu_0 H$ is smaller than the coercivity field, therefore, the driving field does not reverse the magnetization or generate DWs. In order to create DWs by the Oersted field, we injected a current pulse (5 V, 100 ns) through a transverse current line in Fig. 1. As soon as the DWs are created, $\mu_0 H$ pushes the DWs, and then, DW passes through the Hall cross region; the DW arrival time can be detected by monitoring the change in the Hall voltage using an oscilloscope. DW velocity can be calculated from the arrival time and the distance traveled between the writing line and the Hall bar (400 μm). To obtain a sufficiently high signal-to-noise ratio, we averaged the data from 10 repeated measurements. Low temperature probe station was employed to control the wide range of temperature.

We investigated magnetic field-driven DW velocity $v$ for various temperature range. Figure 2(a) shows $\mu_0 H$ dependence of $v$ at different temperatures $T$. For each $T$, different DW dynamic regimes are indicated in Fig. 2(a) from the creep regime (very low magnetic field) to the flow regime (high magnetic field). Please note that a shift of the curve of $(v, \mu_0 H)$ towards the high-field region is shown as $T$ decreases. In order to quantitatively discuss this behavior, we investigated the depinning field $\mu_0 H_{\text{dep}}$ (see the purple arrow in Fig. 2(a)) with respect to $T$ as shown in Fig. 2(b). This figure clearly shows that $\mu_0 H_{\text{dep}}$ monotonically decreases as $T$ increases. From this result, we experimentally found that the temperature plays a decisive role in $\mu_0 H_{\text{dep}}$.

To understand why $\mu_0 H_{\text{dep}}$ exhibits this behavior of $(\mu_0 H_{\text{dep}}, T)$, we examined $T$-dependent magnetic properties. Figures 3(a) and (b) show $T$ dependence of the saturation magnetization $M_S$ and the magnetic anisotropy field $\mu_0 H_K$, respectively. Inset of Fig. 3(b) shows the magnetization $M$ normalized by $M_S$ as a function of in-plane magnetic field $\mu_0 H_{\text{in}}$ for each $T$. $M_S$ exhibits a linear dependence on $T$, whereas $\mu_0 H_K$ is constant over ranges of $T$ (see the red line in Fig. 3(b)). This tendency of $M_S$ with respect to $T$ is consistent with the results given elsewhere [32, 33], because $M_S$ in ferrimagnets is close to zero in the vicinity of the magnetization compensation temperature [30–33]. Please Note that $(1/M_S, T)$ is similar to the tendency of $(\mu_0 H_{\text{dep}}, T)$. From this result, we experimentally found that there exists a correlation between $\mu_0 H_{\text{dep}}$ and $1/M_S$.

To understand the correlation between $\mu_0 H_{\text{dep}}$ and $M_S$, we adopt here the theory of creep-scaling criticality [1, 3, 27]. According to Ref. [1], $\mu_0 H_{\text{dep}}$ is defined as $\mu_0 H_{\text{dep}} = (\varepsilon_{\text{el}} \xi)/(M_S t L_C^2)$, where $\varepsilon_{\text{el}}$ is the DW energy density per unit length, $\xi$ is the characteristic length of the disorder potential, $t$ is the thickness of the ferromagnetic layer, and $L_C$ is the characteristic collective pinning length (called the Larkin-Ovchinikov length). These parameters are further related as $\varepsilon_{\text{el}} = 4t\sqrt{AK_{\text{eff}}}$ and $L_C = [(\varepsilon_{\text{el}}^2 \xi)/(f_{\text{pin}}^2 n_i)]^{1/3}$, where $A$ is the exchange stiffness constant, $K_{\text{eff}}$ is the effective magnetic anisotropy energy, $f_{\text{pin}}$ is the local pinning force, and $n_i$ is the surface density of the pinning centers. By summing up all these parameters, we can finally obtain the following relationship.

$$\mu_0 H_{\text{dep}} = \beta \sqrt{\mu_0 H_K/M_S}. \qquad (1)$$

$\beta$ is defined by

$$\beta \equiv (1/4\sqrt{2})[(\xi^{1/2} \Delta\sigma^2 n_i)/(A^{5/4})]^{2/3}, \qquad (2)$$

where $\Delta\sigma$ is the difference of the Bloch-wall energy density because of the local thickness. Note that $\beta$ in Eq. (2) is independent of $M_S$, $H_K$, and $t$. Therefore, the creep-scaling theory explains the relation between $\mu_0 H_{\text{dep}}$ and $\sqrt{\mu_0 H_K/M_S}$.

Figure 4 shows $\mu_0 H_{\text{dep}}$ as a function of $\sqrt{\mu_0 H_K/M_S}$ using data from Fig. 2(b) and Fig. 3. The red line represents the proportionality given by Eq. (1). Here, the finite y-intercept to the abscissa can be possibly ascribed to changing magnetic properties during the pattering process. From the Ref. [33], the magnetization compensation temperature $T_M$ of the film does not match $T_M$ of the device because of the loss of the Gd moment during the device patterning process. Although this means that $M_S$ of the device is different from $M_S$ of the film for each temperature, $T$ dependence of $M_S$ is the same for the patterned device and the film by shifting $T$ vs. $M_S$ curve [33]. Therefore, we conclude that the linearity between $\mu_0 H_{\text{dep}}$ and $\sqrt{\mu_0 H_K/M_S}$ can be understood on the basis of Eq. (1).

For confirmation of the generality, we examined $\mu_0 H_{\text{dep}}$, $M_S$, and $\mu_0 H_K$ for various GdFeCo samples. We prepared a series of SiN/GdFeCo/SiN and SiN/GdFeCo/Pt/SiN films with various thickness of GdFeCo layers. Table I lists the detailed sample structures. Here, we define that SiN/GdFeCo/SiN samples are referred to as the series I and SiN/GdFeCo/Pt/SiN samples as the series II, respectively. Note that $\mu_0 H_{\text{dep}}$ is independent on $t$ from the Eq. (1). Accordingly, if we investigate $\mu_0 H_{\text{dep}}$ with respect to $t$, we expect that $\mu_0 H_{\text{dep}}$ exhibits the linear dependence on $\sqrt{\mu_0 H_K/M_S}$. Figure 5 shows $\mu_0 H_{\text{dep}}$ as a function of $\sqrt{\mu_0 H_K/M_S}$ for all samples listed in Table I. This relationship is clearly linear. The color lines are the best linear fit. Notably, $\beta$ in series I is larger than that in series II. This result implies that $\beta$ in Eq. (2) is sensitive to the buffer layer which might be attributable to the changing of $\xi$, $\Delta\sigma$, $n_i$, and $A$. Even though $\beta$ in Eq. (2) is strongly depend on the buffer layer, all values $(\mu_0 H_{\text{dep}}, \sqrt{\mu_0 H_K/M_S})$ lie a single curve with a constant slope within the same buffer layer.

Therefore, the results prove the validity of the generality of $(\mu_0 H_{\text{dep}}, \sqrt{\mu_0 H_{\text{K}}/M_{\text{S}}})$.

We observed that the depinning field is sensitive to the temperature. Based on the quantitative analysis, it was revealed that the primary origin of this dependence can be attributed to the variation of the saturation magnetization and magnetic anisotropy field. Our results offer a theoretical description of the relationship between depinning and magnetic properties, and provide a deep understanding of the depinning field in DW motion. This finding helps for an optimal design rule about pinning for better performance of DW-based devices.

Table I. Summary of the sample structures

|  | **Sample Structures** |
|---|---|
| **Sample I** | 5-nm SiN/10-nm $Gd_{25.0}Fe_{65.6}Co_{9.4}$/100-nm SiN/Si substrate |
| **Sample II** | 5-nm SiN/20-nm $Gd_{23.5}Fe_{66.9}Co_{9.6}$/100-nm SiN/Si substrate |
| **Sample III** | 5-nm SiN/30-nm $Gd_{23.5}Fe_{66.9}Co_{9.6}$/100-nm SiN/Si substrate |
| **Sample IV** | 5-nm SiN/10-nm $Gd_{25.0}Fe_{65.6}Co_{9.4}$/5-nm Pt/100-nm SiN/Si substrate |
| **Sample V** | 5-nm SiN/20-nm $Gd_{23.5}Fe_{66.9}Co_{9.6}$/5-nm Pt/100-nm SiN/Si substrate |
| **Sample VI** | 5-nm SiN/30-nm $Gd_{23.5}Fe_{66.9}Co_{9.6}$/5-nm Pt/100-nm SiN/Si substrate |

**Figure Captions**

**Figure 1** Schematic illustration of the device.

**Figure 2 (a)** DW velocity $v$ as a function of the out-of-plane magnetic field $\mu_0 H$ for various temperature $T$. The purple arrow indicates the depinning field $\mu_0 H_{\text{dep}}$. **(b)** $\mu_0 H_{\text{dep}}$ as a function of $T$.

**Figure 3 (a)** The saturation magnetization $M_S$ as a function of $T$. **(b)** The magnetic anisotropy field $\mu_0 H_K$ as a function of $T$. The red line indicates constant value.

**Figure 4** $\mu_0 H_{\text{dep}}$ as a function of $\sqrt{\mu_0 H_K/M_S}$. The red line is the best linear fit.

**Figure 5** $\mu_0 H_{\text{dep}}$ as a function of $\sqrt{\mu_0 H_K/M_S}$ for series I and series II samples. The red and blue lines are the best linear fit.

**Acknowledgements**

This work was supported by JSPS KAKENHI (Grant Numbers 15H05702, 26103002, and 26103004). Collaborative Research Program of the Institute for Chemical Research, Kyoto University, and R & D project for ICT Key Technology of MEXT from the Japan Society for the Promotion of Science (JSPS). This work was partly supported by The Cooperative Research Project Program of the Research Institute of Electrical Communication, Tohoku University. D.H.K. was supported as an Overseas Researcher under Postdoctoral Fellowship of JSPS (Grant Number P16314).

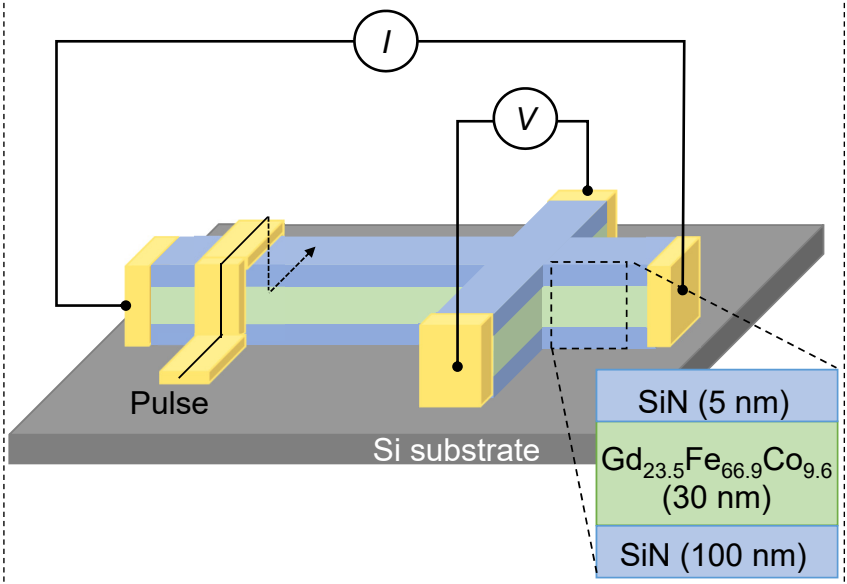

Figure 1

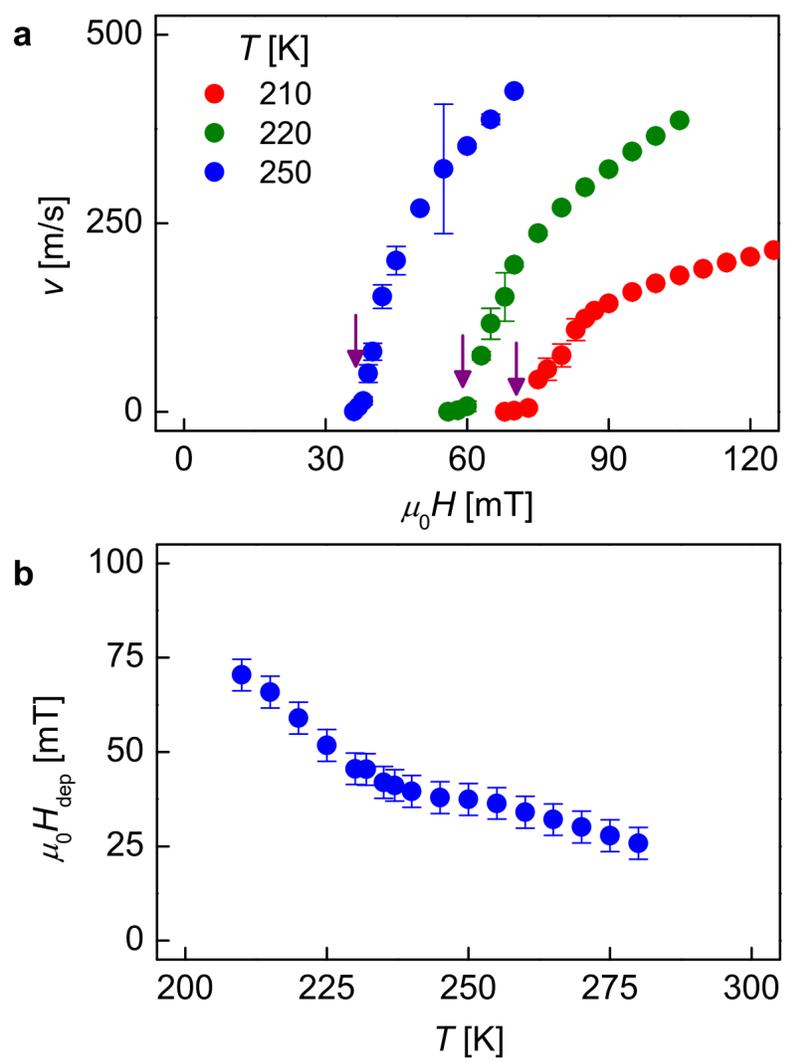

Figure 2

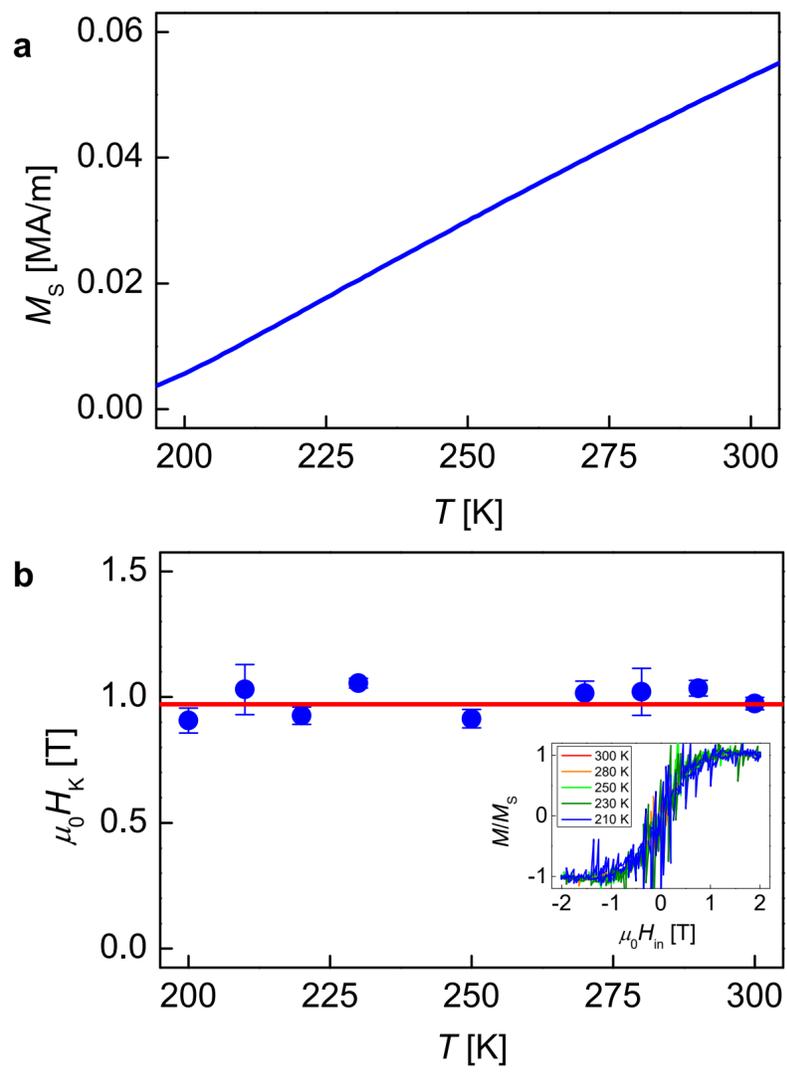

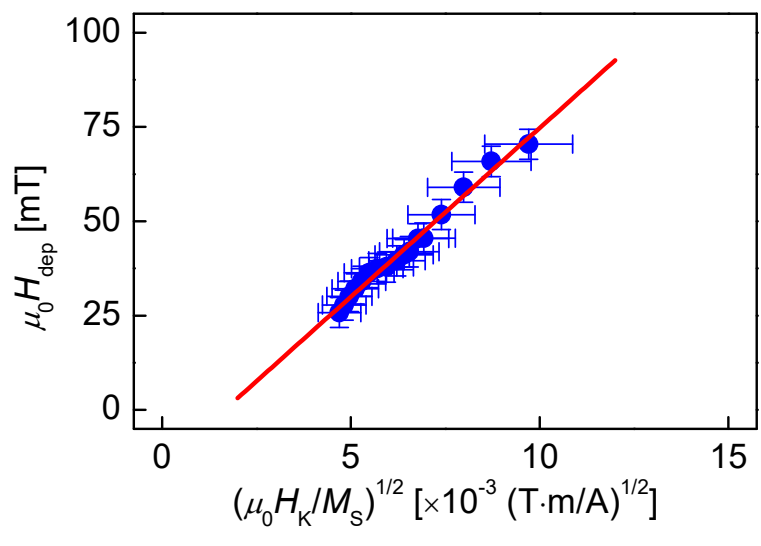

Figure 4

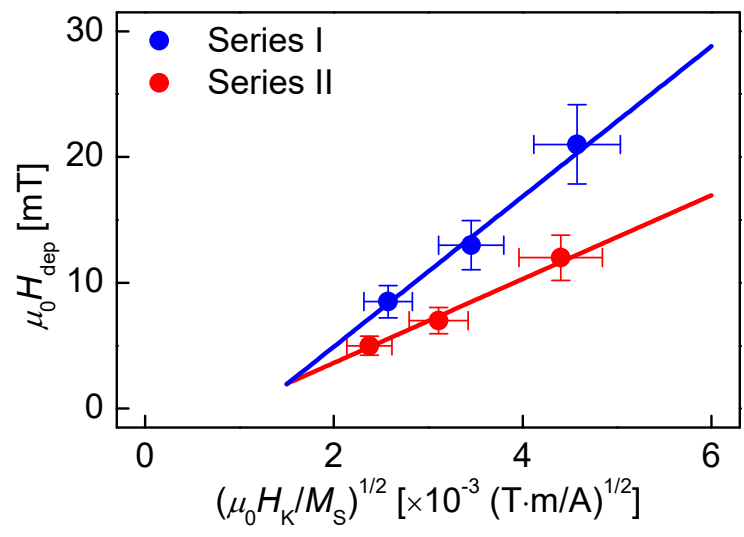

Figure 5